\documentclass[11pt]{article}
 
 \usepackage{amsfonts,amssymb,amsmath}
\usepackage[latin1]{inputenc}

\newcommand{\sect}[1]{\setcounter{equation}{0}\section{#1}}
\newcommand{\subsect}[1]{\subsection{#1}}

\def\be{\begin{equation}}
\def\ee{\end{equation}}
\def\bea{\begin{eqnarray}}
\def\eea{\end{eqnarray}}
\def\ba{\begin{align}}
\def\ea{\end{align}}

\def\JJ{J}

\def\>#1{{\bf #1}}

\def\1{\'{\i}}
\def\R{{\rm I\kern-.2em R}}

  \def\adsw{AdS$_\omega$\ }
\def\k{\omega}
\def\som{\sqrt{\omega}}
\def\sq3{\sqrt{3}}

\begin{document}

\

\bigskip\bigskip

\begin{center}
 {\Large \bf  
Towards (3+1) gravity through Drinfel'd doubles\\[6pt] with cosmological constant
}

\end{center}

 \medskip

\begin{center}

{\sc Angel  Ballesteros, Francisco J. Herranz  and
Pedro Naranjo}

{Departamento de F\1sica, Universidad de Burgos, 
E-09001 Burgos, Spain}

e-mail: {angelb@ubu.es, fjherranz@ubu.es,   pnaranjo@ubu.es}

\end{center}

\begin{abstract}
We present the generalisation to (3+1) dimensions of a quantum deformation of the (2+1) (Anti)-de Sitter and Poincar\'e Lie algebras that is compatible with the conditions imposed by the Chern--Simons formulation of (2+1) gravity. Since such compatibility is automatically fulfilled by deformations coming from Drinfel'd double structures, we believe said structures are worth being analysed also in the (3+1) scenario as a possible guiding 
principle towards the description of (3+1) gravity. To this aim, a canonical classical $r$-matrix arising from a Drinfel'd double structure for the three (3+1) Lorentzian algebras is obtained. This $r$-matrix turns out to be a twisted version of the one corresponding to the (3+1) $\kappa$-deformation, and the main properties of its associated noncommutative spacetime are analysed. In particular, it is shown that this new quantum spacetime is not isomorphic to the $\kappa$-Minkowski one, and that the isotropy of the quantum space coordinates can be preserved through a suitable change of basis of the quantum algebra generators. Throughout the paper the cosmological constant appears as an explicit parameter, thus allowing the (flat) Poincar\'e limit to be straightforwardly obtained.
\end{abstract}

\medskip 




\sect{Introduction}

The quantisation of gravity remains an unsolved problem in fundamental physics (for a recent, general and very readable account, see~\cite{Kiefer}; for a comprehensive reference, see~\cite{Ashtekar}). In an attempt to render the task more tractable, much work has been done in the (2+1)-dimensional case, as gravitation becomes a topological field theory~\cite{Carlipbook,Carliplrr}. In particular, it is well known that in this case gravity can be described as a gauge theory, referred to as a Chern--Simons (CS) theory~\cite{Achucarro,Witten1}. 

In exploiting the above description as a CS theory, quantum groups \cite{Drib,Drinfelda,CP,majid} have proven themselves very relevant in (2+1) gravity, since within the so-called Fock--Rosly construction the phase space of the theory is given by Poisson--Lie groups, which are just the classical counterpart of quantum groups~\cite{FR,AM,AGS,AS}. Moreover, those Poisson--Lie symmetries coming from Drinfel'd double (DD) structures turn out to be outstanding in this context, since the canonical classical $r$-matrix that arises from a given DD structure automatically satisfies the constraints imposed by the Fock--Rosly construction in order to provide a compatible phase space for (2+1) gravity~\cite{cm2,MN,BHMcqg}. As a consequence of all these considerations, the quantum double is widely believed to play an important role in (2+1) quantum gravity~\cite{Bais, MW, BMS, BatistaMajid, BCatmp, Majidtime, Noui}. In fact, proposals to account for the symmetries of (2+1) quantum gravity in terms of quantum double deformations of the isometry groups of spacetimes have been put forward by realising that the Lie groups underlying classical Lorentzian gravity in (2+1) dimensions can be endowed with certain DD structures, that have been fully classified and constructed~\cite{BHMcqg,BHMplb1,BHMplb2,BHMNsigma,BHNcqg}. In this respect, we stress the need to  
work out this problem for Lorentzian kinematical groups with non-vanishing cosmological constant, since they could be the relevant ones in order to provide models for cosmological and astrophysical consequences of quantum gravity phenomena  (see~\cite{AmelinoLRR} and references therein).

However, if one wishes to carry over the above analysis to the realistic case of (3+1)-gravity, the programme advocated for faces an obstacle: the Fock--Rosly construction links quantum groups with CS gravity, but CS is only defined on odd-dimensional manifolds \cite{Nakahara}. Nevertheless, the fundamental structures used in our approach (DDs   and their associated quantum groups) do admit an extension to (3+1) dimensions, regardless of the particular theory of gravity (if any) they might be linked with. Therefore, the motivation of the results here presented relies on the fact that, as DDs seem to describe the symmetries of  (2+1)-gravity, it is not unreasonable that (3+1)-DDs could be used to suggest symmetry constraints on a novel action principle for (3+1) gravity, hopefully paving the way to 
a more amenable quantisation. Moreover, in such a case, said (3+1)-dimensional generalisations of DD structures should lead to the already known DD structures in (2+1) dimensions through a well-defined dimensional-reduction procedure.

In this paper we present the first---to the best of our knowledge---DD structure for the three Lorentzian Lie algebras in (3+1) dimensions, which is just the generalisation of one of the CS-compatible DD structures that have been recently found in (2+1) dimensions~\cite{BHMcqg}. In our approach, the cosmological constant $\Lambda$ is presented as an explicit parameter that allows a simultaneous description for these three Lie algebras in terms of a one-parametric real Lie algebra that we call AdS$_\omega$ such that $\Lambda\propto -\omega$. In particular, the canonical classical $r$-matrix arising from this DD structure generates a twisted $\kappa$-deformation of  the three Lorentzian Lie algebras in (3+1), whose (2+1) dimensional reduction is the specific Poisson--Lie symmetry that has been fully worked out in~\cite{BHMNsigma} and that, for the  case of vanishing cosmological constant, is the $r$-matrix that generates one of the twisted $\kappa$-deformations of (3+1) Poincar\'e algebra that were studied in~\cite{Zakr, LyakLuki, Dasz1, BoPach}. Moreover, by analysing the first-order term of the quantum deformation provided by the classical $r$-matrix, we will show that the first-order noncommutative spacetime, that would arise after quantisation, can be given a manifestly isotropic structure, which, interestingly, is by no means isomorphic to the $\kappa$-(Anti)-de Sitter spacetime. Therefore, DD structures can also be  used in (3+1) dimensions as a guiding principle in order to select those quantum deformations that (hopefully) could play a role as symmetries in a quantum gravity scenario.

The structure of the paper is as follows. In the next section, the basic definitions of the AdS$_\omega$ Lie algebra and their associated Lorentzian homogeneous spaces will be recalled, together with the (complex) transformation that maps the algebra with $\omega\neq 0$ to $\mathfrak{so}(5)$. In section 3, a DD structure for $\mathfrak{so}(5)$ will be introduced, and the   canonical $r$-matrix that generates an $\mathfrak{so}(5)$ Lie bialgebra structure will be presented. In section 4, the previous complex map will be used to obtain a (real) DD structure for the family of AdS$_\omega$ Lie algebras, and to analyse the properties of the quantum deformation arising from the associated Lie bialgebra structure. In particular, the features of the first-order noncommutative spacetime generated by this quantum deformation will be described, and a change of basis that restore the space isotropy (that seemed to be lost in the initial presentation of the symmetries) is explicitly given. A final section devoted to some comments and open problems closes the paper.


\sect{AdS$_\omega$ in (3+1) dimensions from $\mathfrak{so}(5)$}

Let us consider the (3+1) Lorentzian Lie algebras denoted collectively as the one-parameter family of algebras AdS$_\omega$, where $\k$ is a real parameter. In the   kinematical basis 
$\{P_0,P_a, K_a, J_a\},\ (a=1,2,3)$ of generators of time translation, space translations, boosts and rotations, respectively, 
the commutation rules for the AdS$_\omega$ Lie algebra read
\be
\begin{array}{lll}
[J_a,J_b]=\epsilon_{abc}J_c\,,& \quad [J_a,P_b]=\epsilon_{abc}P_c\,, &\quad
[J_a,K_b]=\epsilon_{abc}K_c\,, \\[2pt]
\displaystyle{
  [K_a,P_0]=P_a  }\,, &\quad\displaystyle{[K_a,P_b]=\delta_{ab} P_0}\,,    &\quad\displaystyle{[K_a,K_b]=-\epsilon_{abc} J_c}\,, 
\\[2pt][P_0,P_a]=\k  K_a\,, &\quad   [P_a,P_b]=-\k \epsilon_{abc}J_c\,, &\quad[P_0,J_a]=0\,.
\end{array}
\label{3mas1}
\ee
Therefore, AdS$_\omega$ contains the  Anti-de Sitter  (AdS) Lie algebra $\mathfrak{so}(3,2)$  when  $\k>0$,  the  de Sitter  (dS) Lie algebra $\mathfrak{so}(4,1)$  for   $\k<0$, and   the  Poincar\'e one  $\mathfrak{iso}(3,1)$ for  $\k=0$.

Notice that  setting $\k=0$ is   equivalent to applying an In\"on\"u--Wigner contraction~\cite{IW} associated with the involutive
automorphism of both $\mathfrak{so}(3,2)$ and $\mathfrak{so}(4,1)$ given by the composition of parity ${\cal P}$ and time-reversal ${\cal T}$, namely
$$
{\cal P}{\cal T}  :  \quad P_0\to -P_0 \quad P_a\to -P_a,\quad  K_a\to K_a,\quad J_{a}\to J_{a}\,.
$$
  This automorphism  gives  rise to the following Cartan
decomposition of AdS$_\omega$:
$$
 {\rm AdS}_\omega  ={\mathfrak{h}}  \oplus  {\mathfrak{p}} ,\qquad 
{\mathfrak{h}  }={\rm Span}\{ K_a,J_a \}\simeq \mathfrak{so}(3,1) ,\qquad
{\mathfrak{p}  }={\rm Span}\{  P_0,P_a \}  ,
$$
so that $\mathfrak{h} $ is the Lorentz subalgebra.
Thus,  the   family   of the  three    (3+1)-dimensional Lorentzian  symmetric homogeneous
spaces  with constant    curvature  is defined by  the quotient ${\mathbf
{AdS}}^{3+1}_\k \equiv {\rm  SO}_{\k}(3,2)/{\rm  SO}(3,1) $, where the Lie groups $H = {\rm  SO}(3,1) $  and ${\rm SO}_{\k}(3,2)$ have  ${\mathfrak{h}} $ and  AdS$_\omega$ as Lie algebras,   respectively. In this framework, the parameter $\k$ is just the constant sectional curvature of the homogeneous space, which is related to the cosmological constant in the form $\k\propto -\Lambda$. Expliclty, ${\mathbf{AdS}}^{3+1}_\k $ comprises the three following Lorentzian spacetimes:

\begin{itemize}
\item $\k>0, \Lambda<0$: AdS spacetime ${\mathbf
{AdS}}^{3+1} \equiv  {\rm SO}(3,2)/{\rm  SO}(3,1)$.

\item $\k<0, \Lambda>0$: dS spacetime ${\mathbf
{dS}}^{3+1} \equiv   {\rm  SO}(4,1)/{\rm SO}(3,1)$.

\item $\k=\Lambda=0$: Minkowski spacetime ${\mathbf
M}^{3+1} \equiv  {\rm  ISO}(3,1)/{\rm  SO}(3,1)$.
\end{itemize}
Consequently, the case $\k=\Lambda=0$ corresponds to the flat or vanishing cosmological constant limit. We emphasise that, throughout the paper, the cosmological constant parameter $\omega$ will be explicitly preserved and all the results presented hereafter will hold for any value of $\omega$.

On the other hand, the $\mathfrak{so}(5)$ Lie algebra is defined by the non-vanishing commutation rules 
\be
[\JJ_{ij}, \JJ_{ik}] = \JJ_{jk}  ,\qquad
[\JJ_{ij}, \JJ_{jk}] = -\JJ_{ik} , \qquad
[\JJ_{ik}, \JJ_{jk}] = \JJ_{ij}   ,\qquad i<j<k ,
  \label{ab}
\ee
where $i, j, k =0,\dots,4$. It turns out that the $\mathfrak{so}(3,2)$ and $\mathfrak{so}(4,1)$ algebras ({\em i.e.}, the AdS$_\omega$ Lie algebra with $\omega\neq 0$) can be obtained as an ``analytic continuation" of $\mathfrak{so}(5)$ by defining the kinematical generators in terms of the angular momenta ones $J_{ij}$ as  
\bea
&& P_0=-\sqrt{\k}\, J_{04},\quad\ P_1={\rm i}\sqrt{\k}\, J_{01},\quad\ P_2={\rm i}\sqrt{\k}\, J_{02},\quad\ P_3={\rm i}\sqrt{\k}\, J_{03},\nonumber\\[2pt]
&&  K_1={\rm i} J_{14},\quad\ K_2={\rm i} J_{24},\quad\  K_3={\rm i} J_{34},\quad\ J_1=J_{23},\quad\ J_2=-J_{13},\quad\ J_3=J_{12} .
\label{map}
\eea
This map expresses nothing but the appropriate change of real form that transforms the Lie brackets (\ref{ab}) into  (\ref{3mas1}), and we will use it in section 4 in order to transform the $\mathfrak{so}(5)$ DD structure into an AdS$_\omega$ one.


\sect{A Drinfel'd double structure for $\mathfrak{so}(5)$}

Let us recall that 
a $2d$-dimensional Lie algebra $\mathfrak{a}$  is endowed with a  (classical) DD  structure   \cite{Drinfelda}   if there exists a basis $\{X_1,\dots,X_d,x^1,\dots,x^d \}$ in which the 
Lie brackets for $\mathfrak{a}$ take
the form
\begin{align}
[X_i,X_j]= c^k_{ij}X_k, \qquad  
[x^i,x^j]= f^{ij}_k x^k, \qquad
[x^i,X_j]= c^i_{jk}x^k- f^{ik}_j X_k  \;.\label{agd}
\end{align}
Hence the two sets of generators $\{X_1,\dots,X_d\}$ and $\{x^1,\dots,x^d \}$ form two Lie subalgebras,  $\mathfrak{g}$ and $\mathfrak{g}^\ast$, with structure constants  $c^k_{ij}$ and $f^{ij}_k$, respectively. In fact, $\mathfrak{g} ^*$  is dual to  $\mathfrak{g} $, that is,
\begin{align}\label{pairdd}
  \langle X_i,X_j\rangle= 0,\qquad \langle x^i,x^j\rangle=0, \qquad
\langle x^i,X_j\rangle= \delta^i_j,\qquad \forall i,j   \, .
\end{align}
The pair ($\mathfrak{g} ,\mathfrak{g}^*$) and the  associated vector space 
$\mathfrak{a}=\mathfrak{g} \oplus \mathfrak{g}^*$ are thus endowed with a Lie algebra
structure by means of the brackets (\ref{agd}), while   the dual relations (\ref{pairdd}) provide 
 an  Ad-invariant quadratic form on the DD Lie algebra $\mathfrak{a}$. 
 
A given $2d$-dimensional Lie algebra can present more than one inequivalent DD structures and, on the other hand, (semi)simple Lie algebras are not---in general---even dimensional. Therefore, the classification of DD structures for a given Lie algebra is a challenging problem, that has been fully solved only in the case of four- and six-dimensional Lie algebras~\cite{gomez,Snobl}. In the case of (semi)simple Lie algebras, a general solution can be found provided each algebra is enlarged with a suitable number of central generators (see~\cite{Enrico, Enrico2}).

We recall that in (2+1) dimensions the classification of DD structures for $\mathfrak{so}(2,2)$ (AdS) and $\mathfrak{so}(3,1)$ (dS) Lie algebras is well known~\cite{Snobl}, and it was used in~\cite{BHMcqg} to construct two essentially different classes of classical $r$-matrices that fulfil the Fock--Rosly constraints imposed by (2+1) gravity. As said earlier, we are unaware of a DD structure having been  obtained in the case of the three (3+1) Lorentzian Lie algebras. Therefore, in the sequel, we present a first DD structure for these 10-dimensional Lie algebras, which can be found by constructing the two dual subalgebras $\mathfrak{g}$ and $\mathfrak{g}^*$  in terms of the generators corresponding to the positive (resp.~negative) roots, together with some linear combination of the generators belonging to the Cartan subalgebra. Although this construction is inspired by  the one presented in~\cite{Enrico, Enrico2}, the result is different, since no central extensions are needed, and the DD structure so obtained can be readily contracted to the Poincar\'e case.

More explicitly, 
let us consider the classical Lie algebra  $\mathfrak{c}_2$ spanned by the  Chevalley generators $\{ h_l, e_{\pm l}\}$ 
with $l=1,2$ and  Lie brackets    given by
  \bea 
&& [h_1,e_{\pm 1}] = \pm e_{\pm 1},     \qquad 
   [h_1,e_{\pm 2}] = \mp e_{\pm 2},\ \, \qquad 
   [e_{+1},e_{-1}] = h_1, \cr
&& [h_2,e_{\pm 1}] = \mp e_{\pm 1},   \qquad 
   [h_2,e_{\pm 2}] = \pm 2 e_{\pm 2}, \qquad 
   [e_{+2},e_{-2}] = h_2,        \nonumber \\ 
&& [h_1,h_2] = [e_{-1},e_{+2}] = [e_{+1},e_{-2}] = 0, \nonumber
\eea 
and supplemented by the four new
generators $e_{\pm 3},\, e_{\pm 4}$ defined by
\be 
[e_{+1},e_{+2}]:=e_{+3},    \qquad
[e_{-2},e_{-1}]:=e_{-3}, \qquad 
[e_{+1},e_{+3}]:=e_{+4},    \qquad
[e_{-3},e_{-1}]:=e_{-4}. \nonumber
\ee 
Hence, the 10 generators $\{ h_l,e_{\pm m}\}$   ($l=1,2$ and $m=1,\dots, 4$)  span 
 the Lie algebra $\mathfrak{so}(5)$. Our aim now  is to     show that  $\mathfrak{so}(5)$  is endowed with a  DD   structure.

Let us denote $f_m\equiv e_{-m}$  and   consider the following linear combination of generators belonging to the Cartan subalgebra:
$$
 e_0 :=\frac{1}{\sqrt{2}}\big((1+\mathrm{i})h_1+\mathrm{i}h_2\big)\,, \qquad  f_0:=\frac{1}{\sqrt{2}}\big((1-\mathrm{i})h_1-\mathrm{i}h_2\big)\,.
$$
Next,  if we perform the identification  
$$
X_i\equiv  e_{+i} ,\qquad  x^i\equiv f_i\equiv  e_{-i} ,\qquad i=0,\dots, 4  ,
$$
it is easily seen that the Lie algebra $\mathfrak{so}(5)$ possesses an underlying DD structure obeying (\ref{agd}), the canonical pairing (\ref{pairdd}) being given by the bilinear form
\be
  \langle e_i,e_j\rangle= 0,\qquad \langle f_i,f_j\rangle=0, \qquad
\langle f_i,e_j\rangle= \delta_{ij},\qquad \forall i,j   \, .
\label{pairdd2}
\ee
In other words,  $\mathfrak{a}\simeq  \mathfrak{so}(5)= \mathfrak{g} \oplus \mathfrak{g}^*$  where   $\mathfrak{g}={\rm Span}\{e_0,\dots ,e_4\}$ and  $\mathfrak{g}^\ast={\rm Span}\{f_0,\dots ,f_4\}$. 

Also, the quadratic Casimir operator for $ \mathfrak{so}(5)$ is obtained from the well known canonical Casimir operator associated with any DD structure:
\begin{align}
C&=\frac12\sum_{i}{(x^i\,X_i+X_i\,x^i)} = \frac12\sum_{i=0}^4{(f_i e_i+ e_i f_i)} \,.
\label{cascas}
\end{align}
 The remarkable point is that, once $\mathfrak{so}(5)$  has been shown to be endowed with a DD structure, a 
(quasitriangular) Lie bialgebra structure $( \mathfrak{so}(5) ,\delta)$   can be directly obtained through the canonical  classical
$r$-matrix given by
\begin{align}
r&=\sum_{i}{x^i\otimes X_i} = \sum_{i=0}^4{f_i\otimes e_i}  \,,
\nonumber
\end{align}
whose  skew-symmetric component 
\begin{align} 
r'&=\frac12 \sum_{i} {x^i\wedge X_i} =\frac 12  \sum_{i=0}^4{f_i\wedge e_i}
\label{rmat}
\end{align}
is related to $r$ through the symmetric element
\be\label{omega}
\Omega=r-r'=\frac12\sum_i{(x^i\otimes X_i+X_i\otimes x^i)} =  \frac 12  \sum_{i=0}^4{ ( f_i\otimes e_i + e_i\otimes f_i ) }\,,
\ee
which is Ad-invariant, as it satisfies 
$ [ Y
\otimes 1+1\otimes Y ,\Omega]=0\,,  \forall Y\in  \mathfrak{so}(5)\,.
$
Then, the Lie bialgebra structure $(\mathfrak{so}(5),\delta)$ generated by this $r$-matrix is given by the cocommutator map
\be
\delta(Y)=[ Y \otimes 1+1\otimes Y ,  r']\,,
\quad
\forall Y\in \mathfrak{so}(5)\,.
\label{rcanon2}
\ee

All the above results can be easily rewritten in terms of the angular momenta generators $J_{ij}$
fulfilling~\eqref{ab}, through the following change of basis:
\begin{align}
e_0&=-\tfrac 1{\sqrt{2} } (  J_{04} - \mathrm{i} J_{13} )  \,, & f_0&= \tfrac 1{\sqrt{2} } ( J_{04} + \mathrm{i} J_{13} )  \,, \nonumber \\[2pt]
e_1&=\tfrac{1}{\sqrt{2}} ( J_{23} +\mathrm{i} J_{12} )\,, & f_1&=-\tfrac{1}{\sqrt{2}} (J_{23} - \mathrm{i} J_{12})\,,\nonumber \\[2pt]
\label{ca} e_2&=\tfrac{1}{2} \bigl(J_{01}-J_{34}  -\mathrm{i} (J_{03}+J_{14}) \bigr)\,, & f_2&=-\tfrac{1}{2} \bigl(J_{01}-J_{34} +\mathrm{i} (J_{03}+ J_{14} )\bigr)\,, 
\nonumber\\[2pt]
e_3&=\tfrac{1}{\sqrt{2}} ( J_{24}+\mathrm{i} J_{02})\,, & f_3&=-\tfrac{1}{\sqrt{2}} ( J_{24} -\mathrm{i} J_{02} )\,,\nonumber \\[2pt]
e_4&=\tfrac{1}{2} \bigl( J_{01} +J_{34}  +\mathrm{i} ( J_{03}- J_{14}  ) \bigr)\,, & f_4&=-\tfrac{1}{2} \bigr(J_{01}+J_{34}-\mathrm{i} (J_{03}- J_{14} )\bigl)\,.\nonumber 
\end{align}
The inverse map reads
\begin{align}
J_{01}&= \tfrac 12   (e_2 - f_2 +e_4 - f_4)   \,, & J_{13}  &= -\tfrac {\rm i}{\sqrt 2} (e_0 + f_0)  \,, \nonumber \\[2pt]
 J_{02}  &=   -\tfrac{\rm i}{\sqrt{2}} (e_3 + f_3) \,, & J_{14}   &=  \tfrac {\rm i}2   (e_2 + f_2 +e_4 + f_4) \,, \nonumber  \\[2pt]
J_{03}&=   \tfrac {\rm i} 2  (e_2 + f_2 -e_4 - f_4) \,, &J_{23} &=    \tfrac 1{\sqrt 2} (e_1 -f_1) \,, \nonumber  \\[2pt]
J_{04} &= -\tfrac {1} {\sqrt 2} (e_0 - f_0) \,, &  J_{24}  &= \tfrac 1{\sqrt{2}} (e_3 -  f_3) \,,\nonumber \\[2pt]
 J_{12} &=  - \tfrac {\rm i}{\sqrt 2}  (e_1 + f_1)\,, &J_{34}  &=    -\tfrac 12  (e_2 - f_2 -e_4 + f_4) \, . \nonumber 
\end{align}
From these relations, the quantum deformation of  $\mathfrak{so}(5)$ associated with the DD structure introduced above is generated by the classical $r$-matrix (\ref{rmat}), which in the new basis (and after introducing a global rescaling factor) becomes 
   \be
 r_{J}  = -  {2}\,{\rm i}\,\xi\, r^\prime = \xi \left(J_{01}\wedge J_{14}+J_{02}\wedge J_{24}+J_{03}\wedge J_{34}+J_{04}\wedge J_{13}+J_{12}\wedge J_{23}\right),
\label{ac}
\ee
$\xi$ being the quantum deformation parameter. The canonical pairing (\ref{pairdd2}), the   Casimir   (\ref{cascas}) and the Ad-invariant  element (\ref{omega}) turn out to be 
\be
  \langle J_{ij},J_{kl} \rangle=-\delta_{ik} \delta_{jl}\,, \qquad 
C=-\frac 12 \sum_{0\le i<j\le 4} J_{ij}^2\,, \qquad \Omega= -\frac 12 \sum_{0\le i<j\le 4} J_{ij}\otimes J_{ij}\,.
\label{xa}
\ee

We stress that the DD classical $r$-matrix $r_{J}$ so obtained (which is a solution of the modified classical Yang--Baxter equation) presents 
an ``hybrid character"~\cite{BHMplb1,Dobrev}, as it is formed by the superposition of a standard or quasitriangular  $r$-matrix  and a nonstandard or triangular one, the latter being, in fact,  a Reshetikhin twist. Namely, $ r_{J}  =r_{\rm {standard}}+r_{\rm {twist}}$, where
$$
 r_{\rm {standard}}=  \xi \left(J_{01}\wedge J_{14}+J_{02}\wedge J_{24}+J_{03}\wedge J_{34}+J_{12}\wedge J_{23}\right)\,, \qquad 
 r_{\rm {twist}}= \xi J_{04}\wedge J_{13}\,.
\qquad 
$$
Recall that $ r_{\rm {standard}}$ is just the $\mathfrak{so}(5)$ classical $r$-matrix worked out in~\cite{LBC}, while $ r_{\rm {twist}}$ provides a trivial solution of the 
classical Yang--Baxter equation, since $[J_{04},J_{13}]=0$.

Finally, the  corresponding  $\mathfrak{so}(5)$ Lie bialgebra generated by the classical $r$-matrix (\ref{ac}) has cocommutators (\ref{rcanon2}) given by
\begin{align}
\delta (J_{04})&=0\,, \qquad \qquad \delta (J_{13})=0\,, \nonumber \\
\delta (J_{12})&=\xi \left(J_{04}\wedge J_{23}+J_{13}\wedge J_{12}\right) , \qquad \delta (J_{23})=\xi \left(J_{12}\wedge J_{04}+J_{13}\wedge J_{23}\right) , \nonumber \\
\delta (J_{01})&=\xi \left(J_{04}\wedge (J_{01}-J_{03})+J_{23}\wedge J_{02}+J_{12}\wedge J_{24}+J_{13}\wedge (J_{34}-J_{14})\right) , \nonumber \\
\delta (J_{02})&=\xi \left(J_{04}\wedge J_{02}+(J_{03}+J_{14})\wedge J_{12}+J_{24}\wedge J_{13}+J_{23}\wedge (J_{34}-J_{01})\right) , \nonumber \\
\delta (J_{03})&=\xi \left(J_{04}\wedge (J_{01}+J_{03})+J_{12}\wedge J_{02}+(J_{14}+J_{34})\wedge J_{13}+J_{24}\wedge J_{23}\right) , \nonumber\\
\delta (J_{14})&=\xi \left(J_{13}\wedge (J_{01}-J_{03})+J_{02}\wedge J_{12}+J_{04}\wedge (J_{14}-J_{34})+J_{23}\wedge J_{24}\right) , \nonumber \\
\delta (J_{24})&=\xi \left(J_{12}\wedge (J_{01}-J_{34})+J_{13}\wedge J_{02}+(J_{03}+J_{14})\wedge J_{23}+J_{04}\wedge J_{24}\right) , \nonumber \\
\delta (J_{34})&=\xi \left(J_{13}\wedge (J_{01}+J_{03})+J_{23}\wedge J_{02}+J_{04}\wedge (J_{14}+J_{34})+J_{12}\wedge J_{24}\right) .\nonumber 
\end{align}
We recall that the map $\delta$ provides the first-order in $\xi$ of the coproduct of the quantum $\mathfrak{so}(5)$ algebra generated by $r_{J}$. Alternatively, the dual map $\delta^\ast$ provides the first-order in the group coordinates of the corresponding $\mathfrak{so}(5)$ quantum group. Under the kinematical realization that we will present in the next section, this first-order will provide the essential information about the noncommutative spacetime generated by this DD structure.


\sect{A quantum AdS$_\omega$  deformation from the Drinfel'd double}

We next use the results obtained in the previous section  for the compact form  $\mathfrak{so}(5)$ in order to construct the corresponding  (first-order)  AdS$_\omega$ deformations and underlying noncommutative spacetimes.
 
The map~\eqref{map} can be applied to the full $\mathfrak{so}(5)$ DD structure in such a manner that  
the classical $r$-matrix for AdS$_\omega$ is obtained from (\ref{ac}) in the form
   \be
 r_\k\equiv \sqrt{\k}\,  r_{J}  =    \xi \left(K_{1}\wedge P_{1}+K_{2}\wedge P_{2}+K_{3}\wedge P_{3}+\sqrt{\k}\, J_{3}\wedge J_{1}+ P_{0}\wedge  J_{2}\right)  ,
\label{acc}
\ee
where a $\sqrt{\k}$ rescaling is needed in order to guarantee a non-vanishing $\k\to 0$ limit, and a real structure for the three Lorentzian symmetries is obtained (this rescaling can be rigorously justified through the theory of Lie bialgebra contractions~\cite{LBC}). Therefore, we get the (3+1) generalisation of the (2+1) classical $r$-matrix 
\be
r=z(K_1\wedge P_1+K_2\wedge P_2) +\theta J\wedge P_0\,,
\label{ca21}
\ee
whose underlying DD structure was found in~\cite{BHMcqg}. We recall that the full (2+1) twisted $\kappa$-AdS$_\omega$ quantum algebra generated by~\eqref{ca21} was constructed in~\cite{BHMNsigma}, and possible (3+1) generalisations of classical $r$-matrices for $\kappa$-deformations with cosmological constant were analysed in~\cite{BHMjpcs}. When comparing~\eqref{acc} and~\eqref{ca21}, we note that the new feature of the (3+1) case is the term $\sqrt{\k}\, J_{3}\wedge J_{1}$. On one hand, this new contribution is cancelled in the $\k\to 0$ Poincar\'e limit, which implies that twisted quantum $\kappa$-Poincar\'e in (2+1) and (3+1) provide similar structures (see~\cite{LyakLuki, Dasz1, BoPach} for the explicit construction of the twist generated on the (3+1) $\kappa$-Poincar\'e algebra by the element $P_{0}\wedge  J_{2}$). On the other hand, when $\k\neq 0$, the term $\sqrt{\k}\, J_{3}\wedge J_{1}$ generates a nontrivial quantum deformation within the rotation subalgebra in the (3+1) case (in the (2+1) setting, this subalgebra is one-dimensional).

Regarding the remaining DD structures, it is straightforward to prove that under~\eqref{map} and after the suitable rescalings that guarantee the existence of the $\k\to 0$ limit, the pairing, Casimir operator and Ad-invariant element (\ref{xa}) read
\begin{align}
\langle P_0,P_0\rangle_\k&=-\omega\,, \qquad \langle P_a,P_b\rangle_\k =\omega \,\delta_{ab}\,,
\qquad \langle K_a,K_b\rangle_\k = \delta_{ab}\, , \qquad        \langle J_a,J_b\rangle_\k=-\delta_{ab}\,,\label{pairing}\\
C_\k&=\k\,C=\frac{1}{2}\Big(\sum _{a=1}^3P_a^2-P_0^2+\omega\sum _{a=1}^3\left(K_a^2-J_a^2\right)\Big)\,, \label{cas31}\\
\Omega_\k &=\k\,\Omega= \frac{1}{2}\Big(\sum _{a=1}^3P_a\otimes P_a-P_0\otimes P_0+\omega\sum _{a=1}^3(K_a\otimes K_a-J_a\otimes J_a)\Big)\,,
\label{casz}
\end{align}
which means that the bilinear form~\eqref{pairing} becomes degenerate in the $\omega\to 0$ Poincar\'e limit, whereas the  Casimir $C_\k$  is always well-defined.

At this point, it is worth recalling that in (2+1) dimensions two quadratic Casimirs and, associated with them,  two inequivalent Ad-invariant symmetric bilinear forms exist for the AdS$_\omega$ algebra (see \cite{Witten1,BHMcqg}). Therefore, two inequivalent pairings exist from the DD viewpoint.  One of them is just the (2+1) analogue of~\eqref{pairing}, while the other bilinear form (which does admit a well-defined $\k\to 0$ limit) is the one commonly used in order to reformulate  (2+1) gravity as a CS gauge theory with the relevant isometry group as gauge group. Nevertheless,  the (2+1)  analogue of~\eqref{pairing}  can be shown to generate a gauge theory with the same equations of motion, but with a different symplectic structure~\cite{cm2,MSquat, etera}.  

  On the contrary, in (3+1) dimensions the AdS$_\omega$ algebra has only one quadratic Casimir~\eqref{cas31} (the other one is a quartic polynomial in the generators) and the only possible bilinear form related to it is just~\eqref{pairing}. Therefore, the study of the phase space associated with the (3+1) DD structure that we have just introduced seems to be pertinent.


\subsect{A new AdS$_\omega$ noncommutative spacetime}

In order to have a more precise idea about the noncommutative AdS$_\omega$ spacetime associated with the quantum deformation generated by the $r$-matrix~\eqref{acc}, 
we can compute the cocommutator map $\delta(Y)=[ Y \otimes 1+1\otimes Y ,  r_\k]$, which becomes
\begin{align}
\delta (P_0)&=0\,, \qquad \qquad \delta (J_2)=0\,, \nonumber \\
\delta (J_1)&=\xi \left(P_0\wedge J_3+\sqrt{\omega} J_1\wedge J_2\right) , \qquad \delta (J_3)=\xi \left(J_1\wedge P_0+
\sqrt{\omega} J_3\wedge J_2\right) , \nonumber \\
\delta (P_1)&=\xi \Big((P_1-P_3)\wedge P_0+\omega \big(J_2\wedge (K_1-K_3)+J_3\wedge K_2\big)+\sqrt{\omega} J_1\wedge P_2\Big) , \nonumber \\
\delta (P_2)&=\xi \Big(P_2\wedge P_0+\omega \big(J_1\wedge K_3+J_2\wedge K_2+K_1\wedge J_3\big)+\sqrt{\omega} \big(P_1\wedge J_1+P_3\wedge J_3\big)\Big) , \nonumber \\
\label{coco} \delta (P_3)&=\xi \Big((P_1+P_3)\wedge P_0+\omega \big(J_2\wedge (K_1+K_3)+K_2\wedge J_1\big)+\sqrt{\omega} J_3\wedge P_2\Big) , \\
\delta (K_1)&=\xi\Big(\big(K_1-K_3\big)\wedge P_0+\big(P_1-P_3\big)\wedge J_2+P_2\wedge J_3+\sqrt{\omega} J_1\wedge K_2\Big) , \nonumber \\
\delta (K_2)&=\xi\Big(K_2\wedge P_0+J_3\wedge P_1+P_2\wedge J_2+P_3\wedge J_1+\som \big(K_1\wedge J_1+K_3\wedge J_3\big)\Big) , \nonumber \\
\delta (K_3)&=\xi\Big(\big(K_1+K_3\big)\wedge P_0+\big(P_1+P_3\big)\wedge J_2+J_1\wedge P_2+\som J_3\wedge K_2\Big) .\nonumber 
\end{align}

As usual, this cocommutator provides the first-order (in the quantum deformation parameter $\xi$) of the quantum deformation of the AdS$_\omega$ algebra that is generated by the DD classical $r$-matrix. Now, by introducing the dual coordinates $\{x^{\mu},k^a,\jmath ^a\}$ such that (note that the pairing here is a different one relating Lie algebra generators to the corresponding AdS$_\omega$ group coordinates around the identity)
$$
 \langle x^{\mu},P_{\nu}   \rangle = \delta ^{\mu}_{\nu}\,, \quad \mu\,,\nu =0,\dots ,3\,, \qquad  \langle k^a,K_b \rangle   =   
\langle \jmath ^a,J_b \rangle   =\delta ^a_b\,,
$$
the (first-order) Poisson--Lie brackets among the 4-dimensional spacetime subalgebra $x^\mu$ are easily read off the dual cocommutator map obtained from~\eqref{coco}:
\begin{align}
\{x^1,x^0\}&=\xi\left(x^1+x^3\right),  \qquad  \{x^2,x^0\}=\xi \,x^2, \qquad  \{x^3,x^0\}=\xi\left(x^3-x^1\right), \nonumber \\
\label{nonAdS}
 \{x^a,x^b\}&=0 ,\qquad  a,b = 1,2,3\,.  
\end{align}
These expressions are just the linearisation of the full Poisson--Lie structure on the AdS$_\omega$ Lie group generated by~\eqref{acc}, that could be explicitly constructed through the corresponding Sklyanin bracket (see~\cite{BHMNsigma} for the full construction in the (2+1) case). As a result, by introducing the quantum group coordinates $\{\hat x^{\mu},\hat k^a,\hat \jmath^a\}$, the noncommutative version of this algebra provides the first-order (in the quantum group coordinates) of the quantum AdS$_\omega$ group generated by the DD $r$-matrix~\eqref{acc}. At this point, we would like to 
emphasise that the cosmological constant $\omega$ will explicitly appear in~\eqref{nonAdS} when higher orders in the group coordinates are taken into account (see~\cite{BHMNsigma} for a detailed discussion, as well as~\cite{BHBruno,Marciano}). 

Some analysis of the (first-order) noncommutative spacetime~\eqref{nonAdS}, which is common for the three (3+1) quantum Lorentzian groups, are in order. Firstly, from the classification of real four-dimensional Lie algebras given in~\cite{invariants}, it can be readily shown that the Lie algebra~\eqref{nonAdS} is isomorphic to the $A_{4,6}^{a,b}$ case with $a=b=1$. On the other hand, if the twist term $\xi\,P_{0}\wedge  J_{2}$ is cancelled in the $r$-matrix~\eqref{acc} (which is tantamount to considering the non-twisted $\kappa$-AdS$_\omega$ deformation in (3+1)), the noncommutative spacetime reads
\begin{equation}
\{x^a,x^0\}=\xi\,x^a\,,   \qquad
\{x^a,x^b\}=0 , 
\label{nonAdSunt} 
\end{equation}
which would be the (non-twisted and first-order) $\kappa$-AdS$_\omega$ spacetime. It is important to stress that~\eqref{nonAdSunt} is not isomorphic to~\eqref{nonAdS} as a Lie algebra, since it corresponds to the different case $A_{4,5}^{a,b}$ with $a=b=1$ in the classification~\cite{invariants}. This means that the existence of an underlying DD structure for the $\kappa$-deformation (that in the (2+1) case is directly related to the compatibility with gravity in a CS setting) implies that a twist has necessarily  to be  included and, as a consequence, that the $\kappa$-noncommutative spacetimes have to be modified in an essential way. We emphasise that this requirement also holds    in the (flat) Poincar\'e case, for which the $\kappa$-Minkowski spacetime~\cite{kMinkowski} has become a keystone in the construction of many different noncommutative models (see, for instance,~\cite{Freidelplb, Pachol} and references therein), as well as in a non-vanishing cosmological constant scenario, for which the 
non-twisted $\kappa$-AdS deformation has already been considered in a quantum gravity context~\cite{starodutsev}.

However, it could be argued that the space isotropy of the noncommutative spacetime~\eqref{nonAdSunt} is broken when the twist $\xi\,P_{0}\wedge  J_{2}$ is considered, since we arrive at~\eqref{nonAdS}, where the $x^2$ coordinate seems to be priviliged from the other two ``quantum space" directions. Nevertheless, this difficulty can be circumvented and the space isotropy can be manifestly recovered in this DD quantum deformation by 
considering the following automorphism  of the \adsw algebra:
\be
\begin{array}{ll}
\displaystyle{\widetilde
Y_1=\frac{1}{\sqrt{6}}\,Y_1 +\frac{1}{\sqrt{3}}\,Y_2+ \frac{1}{\sqrt{2}}\,Y_3},&\qquad\displaystyle{Y_1=
\frac{1}{\sqrt{6}}\left(   \widetilde Y_1+ \widetilde
Y_2  -2 \widetilde  Y_3\right)},\\[8pt]
\displaystyle{\widetilde
Y_2=\frac{1}{\sqrt{6}}\,Y_1+ \frac{1}{\sqrt{3}}\,Y_2 -\frac{1}{\sqrt{2}}\,Y_3},&\qquad\displaystyle{Y_2=
\frac{1}{\sqrt{3}}\left(   \widetilde Y_1+ \widetilde
Y_2+\widetilde  Y_3\right)}, \\[8pt]
\displaystyle{
\widetilde
Y_3=  -\frac{2}{\sqrt{6}}\,Y_1+ \frac{1}{\sqrt{3}}\,Y_2},&\qquad\displaystyle{Y_3=
\frac{1}{\sqrt{2}}\left( \widetilde  Y_1- \widetilde
Y_2\right)},\\[8pt]
\mbox{for}\quad \>Y\in\{\>P,\>K,\>J\},&\qquad \widetilde
P_0=P_0 .
\end{array}
\label{ma} 
\ee
In fact, under~\eqref{ma} the classical $r$-matrix (\ref{acc}) is transformed into
\bea
&& \tilde r_{\k}=\xi\left(\tilde
K_1\wedge \tilde P_1+\tilde K_2\wedge\tilde P_2+\tilde K_3\wedge \tilde P_3 
+\frac{1}{ \sqrt{3} }\, \tilde P_0\wedge \bigl(\tilde J_1+\tilde J_2+\tilde J_3 \bigr) \right. \cr
&& \qquad\qquad  
+ \frac{\sqrt{\k}}{\sqrt{3}} \left. \bigl(     \tilde J_1\wedge \tilde J_2+ \tilde  J_2\wedge \tilde J_3+\tilde J_3\wedge\tilde  J_1\bigr) \right),
 \label{mb} 
\eea
while the   pairing (\ref{pairing}),    Casimir (\ref{cas31})   and  Ad-invariant  element (\ref{casz}) are kept invariant. Indeed, we note that 
in~\eqref{mb} the three spatial directions and the three spatial rotations play exactly the same role, a property that is inherited by the associated cocommutator map, which becomes
\begin{align}
\delta (\tilde P_0)&=0\,, \nonumber \\
\delta (\tilde P_a)&=\xi\left\{\Big(\tilde P_a+\tfrac{1}{\sq3}\big(\tilde P_{a+1}-\tilde P_{a+2}\big)\Big)\wedge\tilde P_0+\tfrac{\som}{\sq3}\Big(
\tilde P_{a+1}\wedge\tilde J_{a+1}+\tilde P_{a+2}\wedge\tilde J_{a+2}\Big.\right. \nonumber \\
 &\qquad \qquad \left.\Big.+\tilde J_a\wedge \big(\tilde P_{a+1}+\tilde P_{a+2}\big)+\som \big(\tilde J_a+\tilde J_{a+1}+\tilde J_{a+2}\big)\wedge\tilde K_a\Big)\right.
 \nonumber \\
&\qquad\qquad \left.\Big. 
 -\omega \big(\tilde K_{a+1}\wedge\tilde J_{a+2}+\tilde J_{a+1}\wedge\tilde K_{a+2}\big)\right\}\,, \nonumber \\
 \delta (\tilde K_a)&= \xi\left\{\Big(\tilde K_a+\tfrac{1}{\sq3}\big(\tilde K_{a+1}-\tilde K_{a+2}\big)\Big)\wedge\tilde P_0+\tfrac{1}{\sq3}\left[\som \Big(
\tilde K_{a+1}\wedge\tilde J_{a+1}+\tilde K_{a+2}\wedge\tilde J_{a+2}\Big.\right.\right. \label{cocosym} \\
 & \left.\left.\Big.+\tilde J_a\wedge \big(\tilde K_{a+1}+\tilde K_{a+2}\big)\Big)+\tilde P_a\wedge\big(\tilde J_a+\tilde J_{a+1}+\tilde J_{a+2}\big)\right]+\tilde P_{a+1}\wedge\tilde J_{a+2}+\tilde J_{a+1}\wedge\tilde P_{a+2}\right\}\,, \nonumber \\
\delta (\tilde J_a)&=\frac{1}{\sq3}\,\xi\Big(\som\,\tilde J_a\wedge\big(\tilde J_{a+1}+\tilde J_{a+2}\big)+\big(\tilde J_{a+1}-\tilde J_{a+2}\big)\wedge\tilde P_0\Big)\,, \nonumber
\end{align} 
where the index $a$ is to be cyclically permuted over the set $\{1,2,3\}$. Finally, by following the same construction of~\eqref{nonAdS} from~\eqref{coco}, the first-order noncommutative spacetime spanned by the dual coordinates of the spacetime subalgebra in \eqref{cocosym} reads
$$
\{x^a,x^0\} = \xi\left(x^a+\tfrac{1}{\sq3}\big(x^{a+2}-x^{a+1}\big)\right)\,, \quad \{x^a,x^b\} = 0\,, \qquad a,b=1,2,3\,.
$$
This Lie algebra, which can indeed be proven to be isomorphic to~\eqref{nonAdS}, has recovered the isotropy among the three spatial directions, as 
 is also the case with the $r$-matrix~\eqref{mb}. Thus, space isotropy is restored in the twisted case with DD structure.


\sect{Concluding remarks and open problems}

Since the role of Poisson--Lie groups as phase space symmetries in the CS approach to (2+1) gravity is well established, one should expect that their quantisation ({\em i.e.}, the corresponding quantum groups) should play a relevant role in (2+1) quantum gravity. At this stage, DD structures have been shown in~\cite{BHMcqg} to be useful in order to select which Poisson--Lie structures on the Lorentzian groups satisfy the conditions imposed by the CS framework and, therefore, which quantum deformations of the AdS$_\omega$ of DD-type should be considered as symmetries of the (2+1) quantum theory, in which the cosmological constant appears as an explicit parameter.

As explained in the Introduction, the (naive) generalisation of this approach to describe (3+1) gravity is not possible, due to the absence of a CS form in (3+1) dimensions. Nevertheless, if we assume that the (2+1) AdS$_\omega$ quantum doubles could be physically meaningful structures, the ``correct" (3+1) quantum symmetries should project themselves down to the (2+1) ones when one of the dimensions is suppressed. Here we have shown that this construction is possible by presenting the (3+1) generalisation of one of the two relevant DD quantum deformations of the AdS$_\omega$ Lie algebra, namely the twisted $\kappa$-AdS$_\omega$ deformation. The (2+1) counterpart of this quantum deformation was fully presented in~\cite{BHMNsigma}, and it turns out that the (3+1) deformation arising from a DD structure mimics the two main features of the (2+1) case. Namely, that the introduction of a twist is essential in order to have an underlying DD structure and the fact that the space isotropy of the associated noncommutative spacetime can be preserved.

Notwithstanding the comment of the paragraph above on the lack of a CS form in (3+1) dimensions, there certainly is a possibility worth exploring. As is 
well known, the Chern--Weil theory enables us to obtain the integral of the second Chern character on a 4-manifold $\mathcal{M}$ by evaluating 
the 
integral of the CS three-form on its boundary $\Gamma = \partial\mathcal{M}$. As this CS three-form is the one used to describe (2+1) gravity as a 
gauge theory, it could also be used to describe (3+1) gravity if the integral evaluation above makes sense. Interestingly, this sort of reasoning 
is well known in spin foam models of quantum gravity (see~\cite{Haggard} and references therein). It would indeed be very interesting to 
elucidate whether this geometrical construction admits an algebraic description along the lines presented in this paper, what would provide some 
hint of possible new action principles for (3+1) gravity and, ultimately, of a more tractable quantisation.  

In addition to the above, a number of natural open problems arise from these results. First of all, the explicit construction of the (3+1) AdS$_\omega$ quantum group, whose first-order has been presented here. To this aim, one could start from the Drinfel'd--Jimbo quantisation of the simple Lie algebra $\mathfrak{c}_2$ given in~\cite{Drinfelda,Jimbo} in order to obtain the (standard) quantum $\mathfrak{so}(5)$ algebra and, afterwards, through the ``analytic continuation" map given by~\eqref{map}, transform it into the  full untwisted AdS$_\omega$ quantum structure (which is the one generated by the clasical $r$-matrix~\eqref{acc} without the $P_{0}\wedge  J_{2}$ term). Finally, the contribution to the quantum deformation coming from such a twist term can be computed by making use of the corresponding twist operator, as it was explicitly shown in~\cite{BHMNsigma} for the (2+1) case. However, the change of basis from the quantum $\mathfrak{c}_2$ generators to the kinematical ones is really cumbersome, and a promising alternative route in order to get the all-orders AdS$_\omega$ deformation consists of implementing the approach presented in~\cite{dualJPA}, which is based on the so-called ``quantum duality principle"~\cite{Drinfelda, STS} and takes directly the DD kinematical $r$-matrix~\eqref{acc} as the first--order initial data. In fact, this method was already used in~\cite{dualJPA} for the $\omega=0$ case (see also~\cite{LyakLuki, Dasz1, BoPach}), and seems to be computationally amenable for the construction of the generic twisted AdS$_\omega$ quantum group~\cite{BHN32}. Armed with this, the role of the cosmological constant for higher orders of the noncommutative spacetime relations should become apparent, and further analyses could be carried out. Among them, the study of the relation between the quantum AdS$_\omega$ Casimirs and the corresponding deformed dispersion relations (see the review~\cite{AmelinoLRR}), the construction of the associated curved momentum spaces  (see, for instance,~\cite{MW, Kowalski, KowalskiNowak, relloc, Banburski}) and the possibility of getting a deeper insight into the physical role of twists in quantum gravity would be worth of a thorough analysis. 

Secondly, another challenging problem deals with the identification of the (3+1) generalisation of the second main DD structure of the (2+1) AdS$_\omega$ Lie algebra that has been analysed in~\cite{BHMplb2} and whose noncommutative spacetime generates an $\mathfrak{so}(2,1)$ Lie algebra, thus being much closer than the $\kappa$-deformations  to the Snyder spacetime~\cite{Snyder}. In this context, a general study of all possible DD structures of AdS$_\omega$ should be undertaken in order to see whether other relevant DD symmetries appear in (3+1) dimensions.


  \subsection*{Acknowledgements}

This work was partially supported by the Spanish Ministerio de Econom\1a y Competitividad     (MINECO) under grant MTM2013-43820-P and   by   Junta de Castilla y Le\'on  under   grant BU278U14.  P.~Naranjo acknowledges a postdoctoral fellowship from Junta de Castilla y Le\'on.


\end{document}